\documentclass{revtex4}

\input epsf
\usepackage{amsmath,amssymb}
\usepackage{graphicx}

\draft

\begin{document}
\titlepage
\title{Inflationary attractor property of phantoms}
\author{Xin-He Meng$^{1,2,3}$ \footnote{xhmeng@phys.nankai.edu.cn}
 \ \ Peng Wang$^1$ \footnote{pewang@eyou.com}
} \affiliation{1.  Department of Physics, Nankai University,
Tianjin 300071, P.R.China \\2. Institute of Theoretical Physics,
CAS, Beijing 100080, P.R.China \\3. Department of Physics,
University of Arizona, Tucson, AZ 85721}

\begin{abstract}
There are some motivations to consider inflation driven by a
phantom field. Before entering into some specific models and
perform data fitting, it is important first investigating some
general features that any viable inflation model should hold. The
inflationary attractor property is an important one of those
features. In this paper we  will show that the inflationary
attractor property still holds for canonical and Born-Infeld
phantom fields in the standard Friedmann-Robertson-Walker
cosmology, however, it does not hold for canonical and Born-Infeld
phantom fields in the Randall -Sundrum II cosmology.
\end{abstract}

\maketitle

\textbf{1. Introduction}

Recent observations do not exclude, and even seem to favor, the
equation of state of the dark energy $\omega<-1$
\cite{Perlmutter}. However, most of the popular candidates for
dark energy, such as quintessence \cite{quin}, K-essence
\cite{kessence} and tachyonic scalar fields described by
Born-Infeld (B-I) action\cite{tachyon}, lead to $\omega>-1$. Thus
the sort of matter or entity with equation of state $\omega<-1$,
called "phantom matter", has received increasingly attentions
recently. Such a field has a very unusual dynamics as it violates
null dominate energy condition (NDEC). In the literature, there
are now mainly two effective descriptions of phantom matter.
First, by opposing the sign of the kinetic term in the Lagrangian
of a canonical scalar field, we will get a phantom field and we
call it "canonical" phantom thereafter \cite{phantom}. The
negative sign of the kinetic term will cause instabilities,
however, it was shown in Ref.\cite{carroll-phantom} that the
instability timescale may be long enough to make it a sensible
effective field theory. Thus it is sensible and interesting to
study their cosmological implications \cite{phantomcos}. Second,
by opposing the sign of the kinetic term in the Lagrangian of a
B-I field, we will get the B-I phantom \cite{Li1,
Odintsov-phantom}. The interesting feature of the B-I phantom is
that it admits a later time attractor solution \cite{Li1} while
the ordinary B-I field does not have this important property
\cite{Li3}.

Thus, it is interesting to consider the scenario of inflation
driven by a phantom field \cite{Li2} and/or the density
perturbation generated by the phantom field \cite{Piao}. If an
ordinary inflation model gives a red spectrum, the corresponding
phantom inflation model gives a blue one, which is an important
character, since current WMAP data gives a blue shift spectrum
$n_s\simeq 1.1$. In ordinary inflation model, only hybrid
inflation can provide blue spectrum. Furthermore, in some more
physical models, phantom inflation appears naturally. For example,
it is well-know that the $R+R^2$ gravity can drive an inflation
without an inflaton \cite{Sta}. Generally, those modified gravity
models with a Lagrangian of the type $L(R)$ have two
\emph{inequivalent} formulations: the metric formulation (second
order formulation) and the Palatini formulation (first order
formulation) \cite{Vollick,Flanagan, Wang-R2, Wang1}. However, due
to an observation of Nima Arkani-Hamed, the Palatini formualton
has fine-tuning problems as an effective quantum field theory.
Specifically, matter loops will give rise to a correction to the
action proportional to the Ricci scalar of the metric
\cite{Flanagan}. In this matter loop corrected version of $R+R^2$
gravity, it was shown in Ref.\cite{Wang-loop} that when expressed
in Einstein frame, it corresponds to a Einstein-Hilbert term
plusing a canonical phantom field. Therefore, in this formulation,
the inflation driven by $R+R^2$ gravity corresponds to a phantom
inflation (It is interesting to contrast this to the fact that in
the "pure" Palatini formulation, the $R+R^2$ gravity can not drive
an inflation \cite{Wang-R2}. It remains an interesting problem
which one of those three formulations of modified gravity is the
physical one or all of them are not). Actually, this is the main
reason motivating us to investigate the properties of phantom
inflation seriously.

If inflation is to be truly predictive, the evolution when the
scalar field is at some given point on the potential has to be
independent of the initial conditions. Otherwise, any result, such
as the amplitude of density perturbations, would depend on the
unknowable initial conditions. However, the scalar wave equation
is a second order equation, implying that $\dot \phi$ can in
principle take on any value anywhere on the potential we may be,
and so there certainly is not a unique solution at each point on
the potential. Inflation can therefore only be predictive if the
solution exhibit an attractor behavior, where the differences
between solutions of different initial conditions rapidly vanish
\cite{Brandenberger} (See Sec.3.7 of Ref.\cite{Liddle} for a
review). The inflationary attractor property of B-I field in
standard FRW cosmology is shown to be held in Ref \cite{Zhang3}.

In this paper, we will investigate the inflationary attractor
properties of the canonical phantom and B-I phantoms in standard
Friedmann-Robertson-Walker (FRW) cosmology and Randall-Sundrum II
(RSII) cosmology \cite{RS}. We consider the later case because
inflation on the brane has now comprised a vast literature and
attracted lots of interests (See Ref.\cite{lr,Lidsey} for a review
and references therein). It has now become a standard activity to
consider any  existing inflation model in RSII cosmology. The
inflationary attractor property was shown to be hold for the
canonical scalar field and B-I fields in RSII cosmology in
Ref.\cite{Zhang1}. We will show that the inflationary property
still holds for canonical and B-I phantom fields in standard FRW
cosmology, however, it does not hold in RSII cosmology.

\textbf{2. Attractor property of canonical phantom}

The canonical phantom field is described by the effective
Lagrangian \cite{phantom}
\begin{equation}
L_{phantom}=\frac{1}{2}(\partial_\mu\phi)^2-V(\phi)\label{1}
\end{equation}
where we use the metric signature $\{-,+,+,+\}$. It is the
negative kinetic energy term that distinguishes the phantom field from
the ordinary scalar fields. In a spatially flat FRW universe model, we
can assume that $\phi$ is spatially homogeneous. The energy
density and the pressure are given by
\begin{equation}
\rho=-\frac{1}{2}\dot\phi^2+V(\phi)\label{2}
\end{equation}
\begin{equation}
p=-\frac{1}{2}\dot\phi^2-V(\phi)\label{3}
\end{equation}

The evolution equation of the field $\phi$ is
\begin{equation}
\ddot\phi+3H\dot\phi-V'(\phi)=0\label{4}
\end{equation}

In analyzing the inflationary attractor property, we will use the
Hamilton-Jacobi (H-J) formulation of the Friedmann equation
\cite{HJ}. In this formulation, we will view the scalar field
$\phi$ as the time variable. This requires that the $\phi$ field
does not change sign during the inflation period. Without loss of generality,
we can choose $\dot\phi>0$ in the following discussions.

\textbf{2.1 Standard FRW cosmology}

For a standard FRW cosmology model contained only canonical phantom
fields, the Friedmann equation is
\begin{equation}
H^2=\frac{\kappa^2}{3}(-\frac{1}{2}\dot\phi^2+V(\phi))\label{5}
\end{equation}
where $H=\dot a/a$ is the Hubble parameter and $\kappa^2=8\pi G$.

Differentiating Eq.(\ref{5}) with respect to $t$ and using
Eq.(\ref{4}) gives
\begin{equation}
H'=\frac{\kappa^2}{2}\dot\phi\label{6}
\end{equation}
Substituting this into Eq.(\ref{5}) will give the H-J formulation
of the Friedmann equation
\begin{equation}
(H')^2+\frac{3\kappa^2}{2}H^2=\frac{\kappa^4}{2}V(\phi)\label{7}
\end{equation}
Eqs.(\ref{6}) and (\ref{7}) are the Hamilton-Jacobi equations,
which are more conveniently to be employed in analyzing the
inflationary attractor behaviors than Eqs.(\ref{4}) and (\ref{5}).
In this formulation, one considers $H(\phi)$, rather than
$V(\phi)$, as the fundamental quantities. If we can solve
$H(\phi)$ from Eq.(\ref{6}), by substituting into Eq.(\ref{7}) we
can immediately obtain $V(\phi)$. Therefore, the Hamilton-Jacobi
formalism is also very useful to obtain a large set of exact
inflationary solution (See Ref.\cite{Zhang1} for some examples)
and put general constraints on the form of the potentials
\cite{Liddle2}.

Supposing $H_0(\phi)$ is any solution to Eq.(\ref{7}), which can
be either inflationary or non-inflationary. We consider a
homogeneous perturbation $\delta H(\phi)$ to this solution; the
attractor property will be satisfied if it becomes smaller as $\phi$
increases. Substituting $H(\phi)=H_0(\phi)+\delta H(\phi)$ into
Eq.(\ref{7}) and linearizing, we find that the perturbation obeys
\begin{equation}
\delta
H'(\phi)=-\frac{3\kappa^2}{2}\frac{H_0(\phi)}{H_0'(\phi)}\delta
H(\phi)\label{8}
\end{equation}
which has the general solution
\begin{equation}
\delta H(\phi)=\delta
H(\phi_i)\exp[-\frac{3\kappa^2}{2}\int^\phi_{\phi_i}\frac{H_0(\phi)}{H_0'(\phi)}d\phi]\label{9}
\end{equation}
where $\delta H(\phi_i)$ is the value at some initial point
$\phi_i$. Since $H_0'$ and $\phi$ have the same sign, if $H_0$ is
an inflationary solution, all linear perturbations damp at least
exponentially. Note that the number of e-foldings is
\begin{equation}
N\equiv\int^{t_{end}}_tHdt=\frac{\kappa^2}{2}\int^{\phi_{end}}_\phi\frac{H(\phi)}{H'(\phi)}d\phi\label{10}
\end{equation}
Then Eq.(\ref{9}) can be written as
\begin{equation}
H(\phi)=H(\phi_i)\exp[-3(N_i-N)]\label{11}
\end{equation}
In this form, if $H_0$ is an inflationary solution, it is obvious
that all linear perturbations approach it at least exponentially
fast as the scalar field rolls. In conclusion, the inflationary
attractor property still holds for canonical phantom field in the
standard FRW cosmology.

\textbf{2.2 Randall-Sundrum II cosmolgy}

For a spatially flat FRW model, the Modified Friedman equation in
Randall-Sundrum II model is \cite{RSMF}
\begin{equation}
H^2=\frac{\kappa^2}{3}(\rho+\frac{\rho^2}{2\lambda})\label{11.1}
\end{equation}
where $\lambda$ is the brane tension. When $\rho\ll\lambda$ (the
low energy limit), this reduces to the standard Friedmann
equation; when $\rho\gg \lambda$ (the high energy limit), this
reduces to
\begin{equation}
H^2=\frac{\kappa^2}{6\lambda}\rho^2\label{11.2}
\end{equation}

Assume that during inflation stage, $V\gg\lambda$, thus the Modified
Friedman equation in the high energy limit is
\begin{equation}
H^2=\frac{\kappa^2}{6\lambda}(-\frac{1}{2}\dot\phi^2+V(\phi))^2\label{12}
\end{equation}

Differentiating Eq.(\ref{12}) with respect to $t$ and using
Eq.(\ref{4}) gives
\begin{equation}
H'(\phi)=\frac{3\kappa}{\sqrt{6\lambda}}H(\phi)\dot\phi\label{13}
\end{equation}
Substituting this into Eq.(\ref{12}) will give the H-J formulation
of the Modified Friedmann equation
\begin{equation}
\frac{\lambda}{3\kappa^2}(\frac{H'(\phi)}{H(\phi)})^2+\frac{\sqrt{6\lambda}}{\kappa}H=V(\phi)\label{14}
\end{equation}

Analogous to the previous section,  substituting
$H(\phi)=H_0(\phi)+\delta H(\phi)$ into Eq.(\ref{14}) and
linearizing, where $H_0(\phi)$ is any solution to Eq.(\ref{7}), we
find that $\delta H(\phi)$ obeys
\begin{equation}
\delta
H'(\phi)=(\frac{H_0'(\phi)}{H_0(\phi)}-\frac{3\sqrt6\kappa}{2\sqrt\lambda}\frac{H_0^2(\phi)}{H_0'(\phi)})\delta
H(\phi)\label{15}
\end{equation}
which has with the general solution form
\begin{equation}
\delta H(\phi)=\delta
H(\phi_i)\exp[\int^\phi_{\phi_i}(\frac{H_0'(\phi)}{H_0(\phi)}
-\frac{3\sqrt6\kappa}{2\sqrt\lambda}\frac{H_0^2(\phi)}{H_0'(\phi)})d\phi]\label{16}
\end{equation}
where $\delta H(\phi_i)$ is the value at some initial point
$\phi_i$. Since $H_0'$ and $d\phi$ have got the same sign, if $H_0$ is
an inflationary solution, the integrand within the exponential
term is positively definite if the condition
\begin{equation}
H_0'(\phi)^2>\frac{3\sqrt6\kappa}{2\sqrt\lambda}H_0(\phi)^4\label{17}
\end{equation}
is satisfied. Using Eqs.(\ref{12}) and (\ref{13}) this condition can
be written as
\begin{equation}
\frac{3}{2}\dot\phi^2>V(\phi)\label{18}
\end{equation}
Thus when the initial value of the $\dot\phi$ satisfies
Eq.(\ref{18}), the perturbations will grow exponentially. In
conclusion, the phantom field in RS II model does not have the
inflationary attractor character.

\textbf{3. Attractor property of Born-Infeld phantom}

The Born-Infeld phantom is described by the effective action
\cite{Li1}
\begin{equation}
L_{BI-phantom}=-V(\phi)\sqrt{1-(\partial_\mu\phi)^2}\label{3.1}
\end{equation}

In a spatially flat FRW universe model, we can assume that the scalar field $\phi$ is
spatially homogeneous. The energy density and the pressure are
given by
\begin{equation}
\rho=\frac{V(\phi)}{\sqrt{1+\dot\phi^2}}\label{3.2}
\end{equation}
\begin{equation}
p=-V(\phi)\sqrt{1+\dot\phi^2}\label{3.3}
\end{equation}

The evolution equation of $\phi$ is
\begin{equation}
\frac{\ddot\phi}{1+\dot\phi^2}+3H\dot\phi-\frac{V'(\phi)}{V(\phi)}=0\label{3.4}
\end{equation}

\textbf{3.1 Standard FRW cosmology}

For a homogenous and isotropic standard FRW cosmology model
the Friedmann equation is taken as
\begin{equation}
H^2=\frac{\kappa^2}{3}\frac{V(\phi)}{\sqrt{1+\dot\phi^2}}\label{3.5}
\end{equation}

Differentiating Eq.(\ref{3.5}) with respect to $t$ and using
Eq.(\ref{3.4}) gives
\begin{equation}
H'=\frac{3}{2}H^2\dot\phi\label{3.6}
\end{equation}
Substituting this into Eq.(\ref{3.5}) will give the H-J
formulation of the Friedmann equation
\begin{equation}
H(\phi)^2\sqrt{1+\frac{4}{9}\frac{H'(\phi)^2}{H(\phi)^4}}=\frac{\kappa^2}{3}V(\phi)\label{3.7}
\end{equation}

Substituting $H(\phi)=H_0(\phi)+\delta H(\phi)$ into
Eq.(\ref{3.7}) and linearizing, where $H_0(\phi)$ is any solution
to Eq.(\ref{3.7}), we find that $\delta H(\phi)$  abides by
\begin{equation}
\delta H'(\phi)=-\frac{9}{2}\frac{H_0(\phi)^3}{H_0'(\phi)}\delta
H(\phi)\label{3.8}
\end{equation}
which has the general solution
\begin{equation}
\delta H(\phi)=\delta
H(\phi_i)\exp[-\frac{9}{2}\int^\phi_{\phi_i}\frac{H_0(\phi)^3}{H_0'(\phi)}d\phi]\label{3.9}
\end{equation}
where $\delta H(\phi_i)$ is the value at some initial point
$\phi_i$. Since $H_0'$ and $\phi$ have possessed the same sign, if $H_0$ is
an inflationary solution, all linear perturbations are damped at least
exponentially. In conclusion, the inflationary attractor property
still holds for the B-I phantom field in standard FRW cosmology.

\textbf{3.2 Randall-Sundrum II cosmolgy}

The Modified Friedman equation in the high energy region can be expressed as
\begin{equation}
H^2=\frac{\kappa^2}{6\lambda}\frac{V(\phi)^2}{1+\dot\phi^2}\label{3.12}
\end{equation}
The solution to this equation is analyzed in Ref.\cite{Li2} and it
was shown analytically that the scale factor $a$ expands
exponentially. Thus Born-Infeld phantom can drive an inflation.

Differentiating Eq.(\ref{3.12}) with respect to $t$ and using
Eq.(\ref{3.4}) gives
\begin{equation}
H'(\phi)=3H^2\dot\phi\label{3.13}
\end{equation}
Substituting this into Eq.(\ref{3.12}) will give the H-J
form of the Modified Friedmann equation
\begin{equation}
\frac{2\lambda}{3\kappa^2}(\frac{H'(\phi)}{H(\phi)})^2+\frac{6\lambda}{\kappa^2}H(\phi)^2=V(\phi)^2\label{3.14}
\end{equation}

Substituting $H(\phi)=H_0(\phi)+\delta H(\phi)$ into
Eq.(\ref{3.14}) and linearizing, where $H_0(\phi)$ is any solution
to Eq.(\ref{7}), we find that $\delta H(\phi)$ obeys
\begin{equation}
\delta
H'(\phi)=(-\frac{9H_0(\phi)^3}{H_0'(\phi)}+\frac{H_0'(\phi)}{H_0(\phi)})\delta
H(\phi)\label{3.15}
\end{equation}
which has the general solution
\begin{equation}
\delta H(\phi)=\delta
H(\phi_i)\exp[\int^\phi_{\phi_i}(-\frac{9H_0(\phi)^3}{H_0'(\phi)}+\frac{H_0'(\phi)}{H_0(\phi)})d\phi]\label{3.16}
\end{equation}
where $\delta H(\phi_i)$ is the value at some initial point
$\phi_i$. Since $H_0'$ and $d\phi$ have the same sign, the
integrand within the exponential term is positive definite if the
condition
\begin{equation}
H_0'(\phi)^2>9H_0(\phi)^4\label{3.17}
\end{equation}
is meet. Using Eqs.(\ref{3.12}) and (\ref{3.13}) this condition
can be written as
\begin{equation}
\dot\phi^2>1\label{3.18}
\end{equation}
Thus when the initial value of the $\dot\phi$ satisfies
Eq.(\ref{3.18}), the perturbations will grow exponentially. Note
that the Lagrangian for a spatially homogenous ordinary B-I field
reads
\begin{equation}
L_{B-I}=-V(\phi)\sqrt{1-\dot\phi^2}\label{3.19}
\end{equation}
Thus, the ordinary B-I field satisfies $\dot\phi^2\leq 1$ \emph{a
prior}. For the phantom B-I field, there is no such natural bound.
And we can see that this will make the B-I phantom not a good
candidate for inflaton.

In conclusion, the B-I phantom in RSII model does not have the
inflationary attractor property.

\textbf{4. Conclusions and discussions}

In this paper, we have shown that the inflationary attractor
property still holds for canonical and B-I phantom fields in
standard FRW cosmology model, however, it does not hold for
canonical and B-I phantom fields in RSII cosmology model. Thus,
subsequently, it is worth considering some specific models of a
phantom inflation in the standard FRW cosmolgy to see whether it
can accommodate the data more easily. However, from our analysis,
considering the phantom inflation in RSII model is not a quite
good choice. This reveals another difference between the standard
FRW cosmology and RSII cosmology models. Presently, there is an
extension of the RSII model by adding a Gauss-Bonnet term in the
bulk action \cite{GB}, it is interesting to see whether the
inclusion of the Gauss-Bonnet term can restore the inflationary
attractor property of canonical and B-I phantoms. And we will find
that, quite interestingly, this is indeed the case \cite{Wang}.

\textbf{Acknowledgements}

We would like to thank D.Lyth, S.D.Odintsov, S.Nojiri and Y.S.Piao for
helpful discussions. This work is partly supported by a China
Doctoral Foundation of National Education Ministry and an ICSC-World
Laboratory  Scholarship.


\begin{thebibliography}{99}
\bibitem{Perlmutter} S. Perlmutter el al. Nature \textbf{404} (2000) 955;
Astroph. J. \textbf{517} (1999) 565; D.N.Spergel et al,
astro-ph/0302207;
\bibitem{quin} B.Ratra and P.J.E.Peebles, Phys.Rev. \textbf{D37}
(1988) 3406; R.R.Caldwell, R.Dave and P.J.Steinhardt,
Phys.Rev.Lett. \textbf{80} (1998) 1582;
\bibitem{kessence} Armendariz-Picon, T.Damour and V.Mukhanov,
Phys.Lett. \textbf{B458} (1999) 209 [hep-th/9904075]; T.Chiba,
T.Okabe and M.Yamaguchi, Phys.Rev. \textbf{D62} (2000) 023511;
\bibitem{tachyon} A.Sen, JHEP \textbf{0204} (2002) 048; ibid, JHEP
\textbf{0207} (2002) 065;
\bibitem{phantom} R.R.Caldwell, Phys.Lett \textbf{B545} (2002)
23; G.W. Gibbons, hep-th/0302199;
\bibitem{carroll-phantom} S.M.Carroll,
M.Hoffman and M.Trodden, Phys.Rev. \textbf{D68} (2003) 023509
[astro-ph/0301273];
\bibitem{phantomcos} S.Nojiri and S.D.Odintsov, Phys.Lett.
\textbf{B565} (2003) 1 [hep-th/0304131]; S.Nojiri and
S.D.Odintsov, Phys.Lett. \textbf{B562} (2003) 147
[hep-th/0303117]; P.Singh, M.Sami and N.Dadhich, Phys.Rev.
\textbf{D68} (2003) 023522 [hep-th/0305110]; L.P.Chimento and
R.Lazkoz, gr-qc/0307111; J.G.Hao, X.Z.Li, Phys. Rev. \textbf{D68}
(2003) 083514 [hep-th/0306033]; M.P.Dabrowski, T.Stachowiak and
M.Szydlowski, hep-th/0307128; J.G.Hao, X.Z.Li, astro-ph/0309746;
J.G.Hao, X.Z.Li, Phys.Rev. \textbf{D67} (2003) 107303
[gr-qc/0302100];
\bibitem{Li1} J.G.Hao and X.Z.Li, Phys.Rev. \textbf{D68} (2003) 043501
[hep-th/0305207];
\bibitem{Odintsov-phantom} S.Nojiri and S.D.Odintsov, Phys.Lett. \textbf{B571} (2003)
1 [hep-th/0306212];
\bibitem{Li3} X.Z.Li, J.G.Hao and D.J.Liu, Chin.Phys.Lett.
\textbf{19} (2002) 1584;
\bibitem{Li2} D.J.Liu and X.Z.Li, Phys.Rev. \textbf{D68} (2003)
067301 [hep-th/0307239];
\bibitem{Piao} Y.S.Piao and E.Zhou, hep-th/0308080;
\bibitem{lr} D.Lyth and A.Riotto, Phys.Rept. \textbf{314} (1999) 1;
\bibitem{Sta} A.A.Starobinsky, Phys.Lett.B \textbf{91} (1980) 99;

\bibitem{Vollick} D. N. Vollick, astro-ph/0306630;
\bibitem{Flanagan} \'{E}.\'{E}.Flanagan, astro-ph/0308111; ibid,
gr-qc/0309015;
\bibitem{Wang-R2} X.H.Meng and P.Wang, astro-ph/0308284;
\bibitem{Wang1} X.H.Meng and P.Wang, Class. and Quant. Grav. \textbf{20} (2003) 4949 [astro-ph/0307354]; ibid,
astro-ph/0308031; ibid, hep-th/0309062;
\bibitem{Wang-loop} X.H.Meng and P.Wang, hep-th/0310038;
\bibitem{HJ} D.S.Salopek and J.R.Bond, Phys.Rev. \textbf{D42}
(1990) 3936; A.G.Muslimov, Class.Quant.Grav. \textbf{7} (1990)
231; J.E.Lidsey, Phys.Lett. \textbf{B273} (1991) 42; A.R.Liddle,
P.Parsons and J.D.Barrow, Phys.Rev. \textbf{D50} (1994) 7222
[astro-ph/9408015];



\bibitem{Brandenberger} D.S.Goldwirth, Phys.Lett. \textbf{B243} (1990) 41; R.Brandenberger, G.Geshnizjani and
S.Watson, Phys.Rev. \textbf{D67} (2003) 123510 [hep-th/0302222];
\bibitem{Liddle} A.R.Lidde and D.H.Lyth, Cosmological Inflation and Large
Scale Structure, Cambrigde University Press, 2000;
\bibitem{Zhang3} Z.K.Guo, Y.S.Piao, R.G.Cai, Y.Z.Zhang, Phys.Rev. \textbf{D68}
(2003) 043508 [hep-ph/0304236];
\bibitem{RS} L.Randall and R.Sundrum, Phys.Rev.Lett. \textbf{83}
(1999) 4690 [hep-th/9906064];
\bibitem{Lidsey} J.E.Lidsey, astro-ph/0305528;
\bibitem{Zhang1} Z.K.Guo, H.S.Zhang and Y.Z.Zhang, hep-ph/0309163;
\bibitem{Liddle2} A.Vallinotto, E.J.Copeland, E.W.Kolb and
A.R.Liddle, astro-ph/0311005;

\bibitem{RSMF} P.Binetruy, C.Deffayet, U.Ellwanger and D.Langlois,
Phys.Lett. \textbf{B477} (2000) 285 [hep-ph/9910219];
\bibitem{GB} J.E.Lidsey and N.J.Nunes, Phys.Rev. \textbf{D67} (2003)
103510 [astro-ph/0303168]; S.Nojiri and S.D.Odintsov, JHEP
\textbf{07} (2000) 049; S.Nojiri, S.D.Odintsov and S.Ogushi,
Int.J.Mod.Phys. \textbf{A16} (2001) 5085; S.Nojiri, S.D.Odintsov
and S.Ogushi, Phys.Rev. \textbf{D65} (2002) 023521; J.E.Lidsey,
S.Nojiri and S.D.Odintsov, JHEP \textbf{06} (2002) 026; J.E.Kim,
B.Kyae and H.M.Lee, Phys.Rev. \textbf{D62} (2000) 045013;
\bibitem{Wang} X.H.Meng and P.Wang, work in preparation;
\end{thebibliography}
\end{document}